\documentclass[conference]{IEEEtran}
\usepackage{cite}

\ifCLASSINFOpdf
  \usepackage[pdftex]{graphicx}
\else
  \usepackage[dvips]{graphicx}
\fi
\usepackage{url}

\begin{document}
%
\title{Fighting Spam by Breaking the Economy of Advertising by Unsolicited Emails}

\author{\IEEEauthorblockN{Alexander Schmidtke, Hans-Joachim Hof}
\IEEEauthorblockA{
Munich IT Security Research Group (MuSe), Department of Computer Science and Mathematics\\
Munich University of Applied Sciences\\
Munich, Germany\\
e-mail: AlexanderSchmidtke@gmx.de, hof@hm.edu}}


%


\maketitle

\begin{abstract}
Unsolicited email (spam) is still a problem for users of the email service. Even though current email anti-spam solutions filter most spam emails, some still are delivered to the inbox of users. A special class of spam emails advertises websites, e.g., online dating sites or online pharmacies. The success rate of this kind of advertising is rather low, however, as sending an email does only involve minimal costs, even a very low success rate results in enough revenue such that this kind of advertising pays off. The anti-spam approach presented in this paper aims on increasing the costs for websites that are advertised by spam emails and on lowering the revenues from spam emails. Costs can be increased for a website by increasing traffic. Revenues can be decreased by making the website slow responding as some business gets lost. To increase costs and decreased revenues a decentralized peer-to-peer coordination mechanism is used to have mail clients to agree on a start date and time for an anti-spam campaign. During a campaign, all clients that received spam emails advertising a website send an opt-out request to this website. A huge number of opt-out requests results in increased traffic to this website and will likely result in a slower responsibility of the website. The coordination mechanism presented in this paper is based on a peer-to-peer mechanisms and a so-called paranoid trust model to avoid manipulation by spammers. A prototype implementation for the Thunderbird email client exist. The anti-spam approach presented in this paper breaks the economy of spam, hence, makes advertising by unsolicited emails unattractive.

\begin{IEEEkeywords}
spam; unsolicited email; peer-to-peer; security (key words)
\end{IEEEkeywords}

\end{abstract}


%
\IEEEpeerreviewmaketitle

\section{Introduction}
According to \cite{S13}, spam accounts for $64.1\%$ of all emails. For this paper, spam refers to an unwanted email and spammer refers to the person sending this unwanted email. According to \cite{SHSV11} spam is used for the follwing purposes: advertise services or products (e.g., pharmaceutical drugs, dating sites, etc.), distribute malware, scam

Anti-spam solutions targeting servers used by spammers (blacklists) are of limited use, as nowadays more than $80\%$ of spam messages are sent with the help of botnets \cite{S11}. Hence, no central server is the source of spam but thousands of regular clients using thousands of legitimate email providers. Email filtering based on the content of spam emails \cite{SDHH98} is an effective method to filter spam mails, however it can not be guaranteed that legitimate emails are not classified as spam. Also, spammers usually optimize their spam emails such that they pass Bayesian filters. In spam mails advertising a website, the URL is hard to obfuscate, because a user must have an easy way to go to the advertised website for the spam mail to be successful. According to \cite{S13}, spam advertising sex sites, dating sites or sites selling pharmaceutical drugs make up for $86.52\%$ of all spam. Hence, an anti-spam approach targeting this class of spam is highly relevant. The approach presented in this paper targets on websites advertised by spam. It aims on both increasing the cost to maintain the website as decreasing the revenue of the website. In \cite{Ilg06}, this is described as one way to stop spam as the success rate of spam is very low. Increasing the costs for the advertised website or decreasing the revenue of the website may lead to spam not being profitable any more. The approach presented in this paper increases the costs and decreases the revenue:
\begin{itemize}
\item Cost increase: clients send opt-out request to websites advertised by spam. These opt-out requests result in additional traffic to the website. If the owner of a website pays for traffic to his site, costs are increased to maintain the website.
\item Revenue decrease: clients coordinate to send opt-out requests at nearly the same time, resulting in the website becoming slow in response. The website may loose some business because customers do not want to wait for the page to load. Sometimes spam websites are hosted on compromised hosts. Increasing the traffic to these hosts results in a higher probability that the administrator of the compromised host notices the infection of the host because legitimate services run slower than usual. It is very likely that the administrator then takes down the unwanted website, reducing the success rate for this website to $0\%$. 
\end{itemize}

Multiple clients that received spam send opt-out requests in a so-called campaign. A campaign is defined by one target URL as well as a start date and time. Opt-out requests are sent after the start date and time. Clients participating in a campaign are called comrades. To participate in a campaign, a client must have received a spam mail advertising a website that leads to the URL of the campaign. Hence, it is ensured that clients only send opt-out requests for spam mails they really received. The result of a campaign is much traffic on a site as well as a slow responsiveness of the site. A coordination algorithm based on a peer-to-peer network is used to let comrades agree on a start date and time for their campaign. The coordination algorithm uses a so-called paranoid trust model to avoid manipulations by spammers. As long as the peer-to-peer network implements a Distributed Hash Table (DHT) offering a PUT method to store information in the DHT and a PULL method to receive information from the DHT, any DHT can be used by the coordination algorithm. It is advised to use an existing DHT resilient to Distributed Denial of Service attacks, e.g., a file sharing peer-to-peer network with a huge user base like eMule \cite{emule} or other resilient peer-to-peer networks like NeighborhoodWatch \cite{BSMGSB09}.

It should be noted that the approach sketched may result in a Distributed Denial of Service attack on websites advertised by spam. It is an open legal question if sending an opt-out request using the proposed system is legal. This question can only be answered individually for each country  and is out of scope of this technical paper. However, the existence of other opt-out anti-spam solutions (see Section \ref{ch:relatedwork} for a thorough discussion on related work) indicates, that countries exist where using the anti-spam solution proposed in this paper is legal. 

Another problem are false positives: a spammer could try to abuse the proposed anti-spam solution by sending spam mails for an innocent website, e.g., to discredit the website of a competitor. A careful examination of emails is necessary to avoid false positives. Giving the user the possibility to check the website the anti-spam solution identified as potential target of a campaign may help to avoid false positives. 

The rest of this paper is structured as follows: Section \ref{ch:relatedwork} discusses related work.  Section \ref{ch:systemdesign} gives an overview of the system architecture. Section \ref{ch:campaigncoordinator} presents the distributed coordination mechanism of the proposed anti-spam solution. Section \ref{ch:trustmodel} introduced a paranoid trust model used for the presented approach. Section \ref{ch:implementation} presents a prototype implementation for Thunderbird. Section \ref{ch:evaluation} evaluates the anti-spam approach presented in this paper under different attacks. Section \ref{ch:conclusion} concludes the paper and gives an outlook on future work.

\section{Related Work}\label{ch:relatedwork}
This section presents related work on fighting spam by increasing the cost of spam and related work on using a DHT to fight spam. 

Hashcash \cite{B97}\cite{B02} is used to target Denial of Service attacks. Hashcash is a challenge-response approach. In \cite{B97}\cite{B02}, hashcash is used to increase the cost of sending email. To do so, when an email clients delivers an email to a mail server, the email server does not accept the email at once. Instead, the mail server provides a challenge to the mail client. The challenge requires some computational effort to be solved. The solution of the challenge is sent together with the mail. The mail server only accepts mails that include a solved challenge. Hence, the solved challenge can be seen as a virtual stamp. With this system, sending a single mail is still fast, but sending thousands of messages is significantly slowed, resulting in a lower spam per time rate  increasing the cost of spam. However, there are some disadvantages:
\begin{itemize}
\item Hashcash helps email providers to throttle down the number of mails its users can sent. This helps an email provider to avoid abuse of its service. However, spammers often have an email provider of their own or directly deliver the message. Also, service for users of the email provider is limited.
\item Hashcash only avoids sending a huge bulk of emails. According to \cite{S11}, more than $80\%$ of spam is sent with the help of a botnet. Hence, each zombie of a botnet has a much smaller bulk of emails to sent. Hashcash is not very effective to avoid spam sent by a botnet.
\end{itemize}

Some anti-spam approaches already use a DHT to fight spam, e.g., NeighborhoodWatch \cite{BSMGSB09}, a blacklist of IP addresses of known spammers. Although NeighborhoodWatch prevents Distributed Denial of Service attacks on the blacklist, it does not increase costs for spammers, hence, does not target the business model of spammers. However, the DHT used by NeighborhoodWatch can be used for the approach presented in this paper.

The startup company Blue Frog \cite{L06} offered a software that let email users send coordinated opt-out requests for spam they received similar to the approach presented in this paper. Blue Frog was very successful and forced six of the top ten bulk email groups to cooperate with them to remove users of the Blue Frog software from their email lists \cite{L06}. Hence, Blue Frog showed that a coordinated opt-out approach to fight spam can be very successful. Lycos \cite{BBC04} offered a similar approach with its "Make Love not Spam" campaign. However, both Lycos and Blue Frog used a central coordination for the period of time the clients should send opt-out requests. This central coordination is vulnerable to attacks. In the case of Blue Frog, the central coordination mechanism was brought down by a massive Denial of Service attack. Another disadvantage of Blue Frog was the way users submitted received spam mails: users were asked to forward received spam mails. This resulted in users being blacklisted by mail providers that scan outgoing mails for spam mails. The approach presented in this paper follows the idea of coordinated opt-out requests. The success of Blue Frog and Lycos showed that the traffic generated by sending coordinated opt-out requests has a significant impact on the business model of spammers and websites advertised by spam. However, the approach presented in this paper tries to avoids the disadvantages of the solutions of Blue Frog and Lycos: no central coordination is used and users do not have to forward spam mails.

\section{System Design}\label{ch:systemdesign}
This section gives an overview of the proposed anti-spam solution. The following sections focus on one part of the system: the Campaign Coordinator. The Campaign Coordinator is the main difference to existing anti-spam solutions as discussed in Section \ref{ch:relatedwork}. 

The proposed anti-spam solution has four components: Spam Classifier, Target Evaluator, Campaign Coordinator, and Opt-Out Module

The \textit{Spam Classifier} decides which emails are spam and which are not. One possible implementation includes a user interaction that allows a user to express which emails are spam from his point of view. Involving the user for spam classification has the benefit, that only the user can decide what "unwanted" mail means for him. Also, involving the user may help to avoid false positives. If not enough users decide that an email is spam, no campaign will take place. The Spam Classifier passes spam mails to the Target Evaluator.
The \textit{Target Evaluator} extracts one or more URLs from a spam email. As a user must have a possibility to go to the website advertised by a spam mail, it is assumed that the identification of the URL of the website is possible. The Target Evaluator outputs URLs of the advertised website. A careful evaluation of the mail content is necessary because of the following reasons:
\begin{itemize}
\item  Spam mails may use redirector services like TinyURL \cite{tinyurl} for advertised website. The URLs of the redirection service may even be different for each spam mail. While getting a redirection URL already increases the cost of sending spam, this may be efficient to avoid spam campaigns because all URLs are different. To avoid this attack, the Target Evaluator looks for the use of redirection services.
\item URLs in spam mails may contain parameters that identify emails of users. Hence, going to this URL may verify the email address of a receiver, potentially resulting in getting even more spam. Also, parameters of a URL may be different for each user. Hence, the Target Evaluator needs to strip parameters from URLs.
\item Some email providers add footers to outgoing mail advertising their service. This footers usually include the URL of the mail server providers website, e.g., "This mail was sent by www.gmx.de". The Target Evaluator has a whitelist of URLs not to attack to avoids that such legitimate URLs become targets of anti-spam campaigns.
\item An attacker could try to abuse the anti-spam solution by sending URLs of innocent websites in spam mails to provoke an attack on the innocent website. 
\end{itemize}
It may be a good idea to involve the user in identifying the URLs to avoid false positives. The user decides on the websites to attack. The Target Evaluator is out of scope of this paper. The Target Evaluator passes one ore more URLs to the Campaign Coordinator.

The \textit{Campaign Coordinator} is the heart of the anti-spam approach presented in this paper. The Campaign Coordinator uses a peer-to-peer network to identify other clients that also received a spam mail advertising the same URL. All clients that received this spam mail decide on a date and time when to start a campaign against the website with the given URL. The Campaign Coordinator is described in more detail in \ref{ch:campaigncoordinator}.

The \textit{Opt-Out Module} is invoked at the date and time when a campaign starts. The module starts to send opt-out requests to the website. Please note that the Opt-Out Module does not depend on the availability of an opt-out link in the spam mail. One possible implementation of the attack module scans the website at the given URL for images and sends the opt-out request as a parameter in a GET request to all images. This opt-out request will show up in the log of the web server. Sending opt-out requests in this way can not even be stopped by using a captcha on the website to avoid automated opt-out requests. The opt-out requests result in traffic to the site, and the coordinated start of the campaign may result in a slower response of the website due to many requests.  The Opt-Out Module is out of scope of the paper, successful opt-out anti-spam solutions like those discussed in Section \ref{ch:relatedwork} show that enough traffic can be generated to have an impact on the business model of websites advertised by spam.

%


The rest of this paper focuses on the central component of the system: the Campaign Coordinator. The Campaign Coordinator is the main difference to existing anti-spam solutions as discussed in Section \ref{ch:relatedwork}.

\section{Design of the Campaign Coordinator}\label{ch:campaigncoordinator}
The Campaign Coordinator is the heart of the anti-spam approach presented in this paper. A campaign is identified by exactly one URL. If the Target Evaluator extracts more than one URL from a spam mail, it invokes the Campaign Coordinator once for each URL.

The Campaign Coordinator uses a DHT like Kademlia \cite{MaMa02} or Chord \cite{SMKKB01}. Only PUT and GET options of these DHTs are used, hence, it is not necessary to start a new DHT but an existing DHT can be used. Many file sharing systems are based on DHTs. One of those could be used for the approach presented here. Reusing an existing DHT has the benefit of using a large peer-to-peer network that is not as vulnerable to attacks (e.g. Sybil attack) as a small network. Reusing an existing network may also make it easier to join the network \cite{CoHo07}.

\begin{figure}
 \centering
  \includegraphics[width=2.5in]{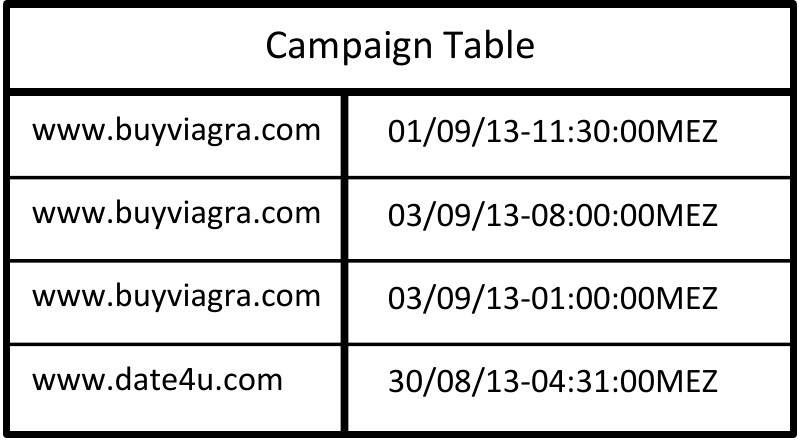}
  \caption{Example for a campaign table holding two campaigns. }\label{fig:campaigntable}
\end{figure}
\begin{figure}
 \centering
  \includegraphics[width=2.5in]{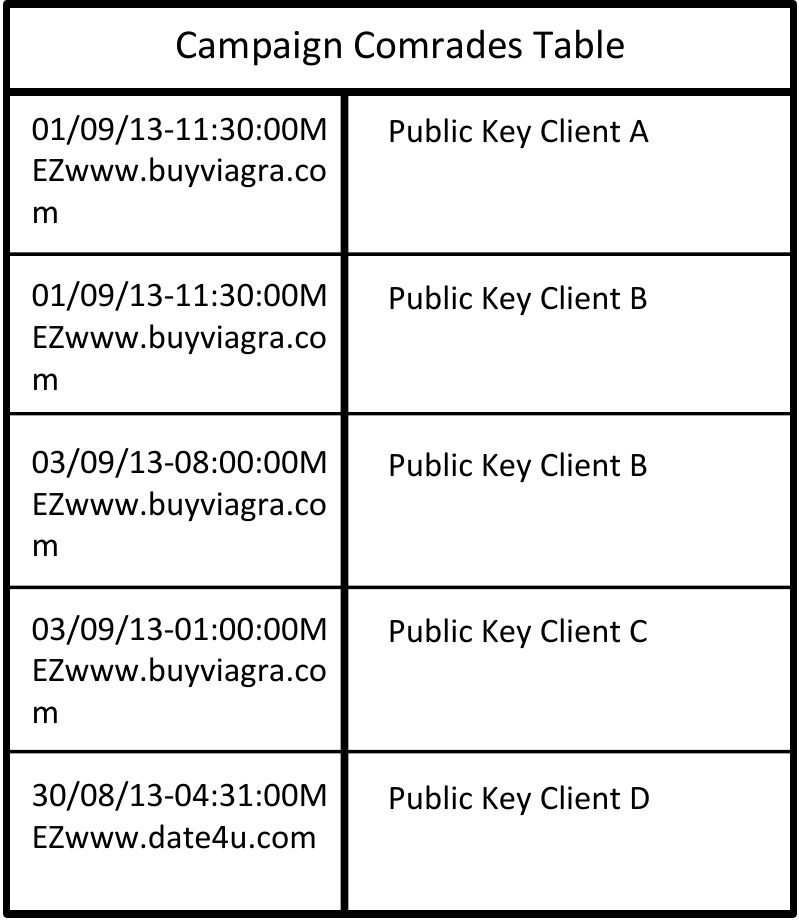}
  \caption{Example for a campaign comrades table for Figure \ref{fig:campaigntable}.}\label{fig:campaigncomradestable}
\end{figure}
\begin{figure}
 \centering
  \includegraphics[width=2.5in]{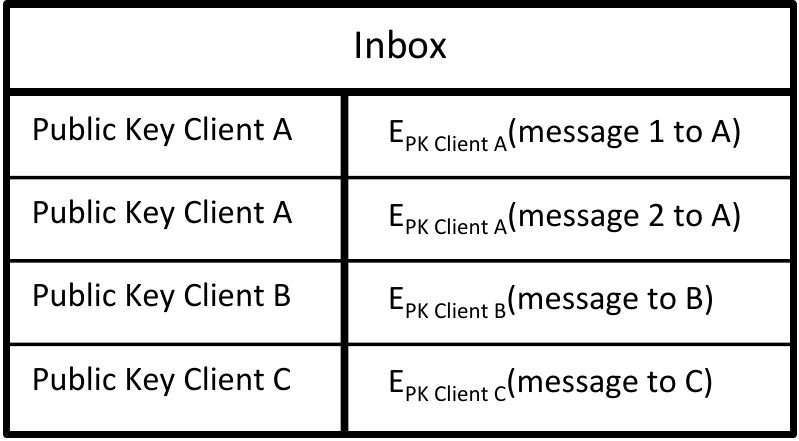}
  \caption{Example for an inbox. }\label{fig:inbox}
\end{figure}

Campaign Coordinators access three different tables stored in the DHT:

\emph{Campaign Table}: the Campaign Table lists available campaigns. A campaign is identified by the URL of the website that is the target of this campaign. The database key to the Campaign Table is the URL, the hash value of the URL is used to access the DHT. The entries in the campaign table list one or more start times and start dates for this campaign. Figure \ref{fig:campaigntable} gives an example of a Campaign Table: one of the campaigns has multiple start dates and times.

\emph{Campaign Comrades Table}: the Campaign Comrades Table lists all the clients that are willing to participate in a given campaign at a given time. Clients participating in a campaign are called comrades in the following. The database key for the Campaign Comrades Table is the start date and time of the campaign concatenated with the URL. The corresponding hash value is used to access the DHT. If multiple start dates and times exist for one campaign, there are multiple database keys for the Campaign Comrades Table.  Figure \ref{fig:campaigncomradestable} gives an example for a Campaign Comrades Table: there is one start date and time with two comrades, the other start dates and times only have one comrade. Client B decided to participate in two campaigns, the other clients only participate in one campaign. It should be noted that the concatenated start date and time and URL are not actually stored in the DHT. Instead, the hash value of the concatenated start date and time and URL is used as access database key to the DHT. Please note that the URLs in this example are only examples - it is not known to the author that any of these URLs have been advertised using spam mails.

\emph{Inbox}: The inbox is used to allow clients to receive encrypted messages. The database key of this table is the public key of the client. The hash value of the public key of the client is used to access the DHT. A message to a client is sent by encrypting the message with the public key of the client and storing in the DHT using the hash value of the database key to access the DHT. A message is received when a client gets all entries in the DHT for the hash value of its private key. Read messages should be removed. Figure \ref{fig:inbox} showns an example of the Inbox table: Client A has two pending messages, Client B and C have one pending message. $E_{PK}(message)$ denotes that $message$ is encrypted using public key $PK$. 

The configuration of the Campaign Coordinator uses the following parameters:
\begin{itemize}
\item Keying Material: a public key together with the associated private key. The public key is used as identity of this client in the Campaign Comrades Table as well as the address of his Inbox in the DHT. 
\item Acceptable Time Interval: The parameters $max\_wait$ and $min\_wait$ are used to decide on valid campaign start dates and times.
\item Minimum Size of Campaign: $min\_comrades$ , the minimum number of comrades in one campaign to participate in this campaign.
\item Minimum Accumulated Trust Values: $min_accummulated_trust$, the minimum accumulated trust value of all comrades in a campaign that is necessary that a Campaign Coordinator takes place in a campaign. See Section \ref{ch:trustmodel} for details of the so-called paranoid trust model used for the proposed anti-spam solution.
\item Usage Time: information on when the client is available and can participate in campaigns. Usage time is either manually defined by the user, or usage of the mail client and uptime of the system are monitored.
\item Trust Database: a database holding trust information about other clients. See Section \ref{ch:trustmodel} for details of the so-called paranoid trust model used for the proposed anti-spam solution.
\end{itemize}

The Campaign Coordinator is invoked by the Target Evaluator. It receives one URL identifying one campaign. After invocation, the Campaign Coordinator follows the following procedure:
\begin{itemize}
\item Step 1: The Campaign Coordinator checks if there is already one or more entries for the given URL in the Campaign Table. It either selects one or more campaign start date and time from the DHT or decides to propose a better suited start date and time to other clients. If the Campaign Coordinator proposes another start date and time it stores the alternative start date and time in the Campaign Table. It uses the hash value of the URL to access the DHT. A campaign start date and time must meet the following conditions to be considered suitable for the client:
\begin{itemize}
\item A campaign start date and time may not be more than $max\_wait$ minutes or less then $min\_wait$ minutes in the future. $max\_wait$ and $min\_wait$ are local settings of the Campaign Coordinator. This setup avoids that a spammer inserts a campaign start date and time into the Campaign Table that lies many years in the future, hence, no attack takes place, or repeatedly adds a campaign start date in the very near future to avoid that many clients participate in a campaign.
\item A campaign start date and time should be in a time period where the computer of the client is typically running. To do so, the Campaign Coordinator considers the typical usage times of the client. For example, if the Campaign Coordinator is implemented as a plugin for an email client, it will prefer the date and times a user typically uses the email client. 
\end{itemize}
A new entry is inserted into the Campaign Table if the Campaign Table has no entry for the URL yet or if the present start dates and times are not suitable. If multiple campaign start dates exist and more than one of them are suitable for the client, the Campaign Coordinator either joins all suitable campaigns or decides to join the campaign with the highest trust value. See Section \ref{ch:trustmodel} for details about the trust model.

\item Step 2: The Campaign Coordinator adds his public key for all selected start dates and times to the associated Campaign Comrades Tables. The database key for a given start time and date in the DHT is the hash value of the campaign start data concatenated with the URL of the campaign. Hence, different start dates result in different database keys. 

\item Step 3: The Campaign Coordinator adds the URL of the campaign to a local database together with the selected start dates and times. The database holds all currently active campaigns for this Campaign Coordinator. The processing of the URL ends.
\end{itemize}

The Campaign Coordinator regularly checks if one of the campaigns in his local database recently started. If so, the client again checks if he wants to take part in the campaign. To do so, it retrieves the information from the Comrades Tables again. At this point in time, the Comrades Table holds a list of all comrades that take place in this campaign. Comrades are identified by their public key. The Campaign Coordinator checks if more than $min\_comrades$ clients participate in the campaign. Only if the number of comrades is high enough, hence, the campaign will involve enough clients to be effective, the client participates in this campaign. Optionally, the client checks if the enough participating comrades are trusted. See Section \ref{ch:trustmodel} for details on the used trust model. If the Campaign Coordinator decides to participate in a campaign, it invokes the Opt-Out Module. The Opt-Out-Module sends the opt-out requests to the URL of the campaign.

It should be noted that all information about campaigns are public. Hence, a spammer may know about a campaign in the future targeting one or more websites advertised by him. The actions of the website administrator may include setting up firewall rules to block access for certain nodes to mitigate the attack as well as increasing server capacity. However, both actions involve additional costs. Hence, the goal of a campaign is still achieved. 

\section{Trust Model}\label{ch:trustmodel}
The Campaign Manager holds a database with trust values for known clients. Clients are identified by their public key. These public keys come from the Campaign Comrade Table, see Section \ref{ch:campaigncoordinator} for details. Several methods are used to establish trust:

\emph{Paranoid Trust Model:} In the paranoid trust model, the client only believes what it sees. During a campaign, a client probes the responsiveness of the URL of the campaign. If the responsiveness decreases significantly the campaign is a success, the campaign manager increases the trust value of all comrades involved in this campaign. A client may even test the responsiveness of a website if it decided not to take part in the associated campaign. This helps to built up trust quickly. If a client detects that several campaigns did not succeed, it resets its trust database. This is an extra countermeasure to avoid manipulation of the trust database.

\emph{Trust by Challenge-Response:} A challenge response-approach is used to establish trust with an unknown client. Trust is only established with clients that participate in one of the own campaigns. A client A wants to establish trust with an unknown client B. To do so, client A generates a challenge that requires some computational effort to solve it (see below for details of the generation of the challenge). Client A stores the response to the challenge together with the public key of client B. Client A generates a message for client B holding Client A's public key as well as the challenge. Client A encrypts the message with client B's public key. The public key of client B usually comes from the Campaign Comrade Table. Client A stores the message in the Inbox of client B. When the Campaign Manager of client B checks its Inbox, it decrypts the message using its private key. Client B solves the challenge. Client B generates a message holding the solution to the challenge as well as the public key of client B. Client B encrypts the message with the public key of client A and stores it in the Inbox of client A. When the Campaign Manager of client A checks its Inbox, it compares the response received to the response stored locally under the given public key. Trust by Challenge-Response makes it harder for an attacker to generate thousands of identities in the DHT. Significant computational resources are needed to generate many identities. Also, many identities do only allow to manipulate campaigns in small boundaries. It cannot be avoided that a campaign takes place. A spammer may have a botnet at hand that offers many computational resources. However, using a botnet to manipulate the proposed anti-spam solution also includes costs. Hence, the goal of the proposed anti-spam solution is still achieved. 

Campaign Coordinators check on a regular basis if there are messages in their inbox. Campaign Coordinators only answer a given maximum of challenges to avoid Denial of Service Attacks.

To generate a challenge for client B, a client A concatenates its public key $PK_{Client B}$ with two random numbers $rand1$ and $rand2$. $rand1$ and $rand2$ are taken from the intervall $[0,max\_rand]$ where $max\_rand$ is a parameter for configuration of the time needed to solve the challenge. A hash value $h$ is calculated using the hash function $hf()$:

\begin{equation}
h=hf(PK_{Client_B}||rand1||rand2)
\end{equation}

The challenge for client B is to find out which value $rand2$ was used in the calculation of $h$. The challenge sent to client B is $rand1$. $PK_{Client A}$ is known from the message as it identifies a client and is needed to send a message to client A.

To solve the challenge, client B follows the following procedure:
\begin{enumerate}
\item Initialize $test$ with 0
\item Calculate $h'=hf(PK_{Client A}||rand1||test)$
\item If $h'!=h$ increase $test$ by 1 and go to step 2
\item $test$ is the solution of the challenge.
\end{enumerate}

The challenge includes the public key of client A to avoid that an attacker forwards received challenges to a legitimate client that solves the challenge and sends it to the attacker that sends the response back to the client. This man-in-the-middle attack is prevented by the inclusion of the senders public key in the challenge.

The trust values from the local trust database of the Campaign Coordinator can be used to decide if a Campaign Coordinator participates in a campaign or not. The Campaign Coordinator only participates if the sum of all trust values of the comrades of this campaign is higher than $min_accummulated_trust$. $min_accummulated_trust$ is a configuration parameter of the campaign manager. To allow for a quick start, $min_accummulated_trust$ starts at a low level and is increased with each successful campaign.

\section{Prototype Implementation}\label{ch:implementation}
The Campaign Coordination described in Section \ref{ch:campaigncoordinator} is agnostic to the underlaying peer-to-peer network. For the implementation on Windows 7 the Campaign Coordinator relies on Overlay Weaver \cite{SYS08}, however changing the DHT only requires minimal changes. Target Evaluator, Campaign Coordinator, and Opt-Out Module are implemented as a program running in the system tray. This program communicates with the Spam Classifier implemented as Addon for the Thunderbird mail client. The Spam Classifier is a simple button for the user to identify spam in this prototype implementation. A more advanced Spam Classifier will be part of further work. The separation of the Spam Classifier from the other components of the anti-spam solution allows an easy adaption of the implementation for different email clients as only the Spam Classifier must be adapted to different mail clients. 

\section{Discussion of Attacks}\label{ch:evaluation}

This chapter evaluates the outcome of attacks on the Campaign Coordinator. 

One possible attack is a \emph{Man-in-the-Middle Attack} on the trust establishment. An adversary can forward challenges received by clients to built trust to other clients to solve it for them. However, this attack is not possible because the public key of the sender of a challenge is included in the challenge itself. 

Another class of attacks are \emph{Denial of Service Attacks}. An adversary can generate many challenges for one client. If this client wants to establish trust, it will solve all the challenges, doing many computations. This attack is not successful because clients only solve challenges sent by clients that participate in one of their campaigns and the number of challenges to solve is limited. 

In a \emph{Time Portal Attack} an adversary tries to move a campaign in the very distant future such that the attack never takes place. To do so, the adversary injects a campaign for a URL into the Campaign Table that has a start date and time that lies years in the future. Clients joining this campaign will never send opt-out requests. In a \emph{Separation Attack} an adversary tries to reduce the number of comrades for a URL. This is done by repeatedly injecting campaigns for the URL with a very early start date and time. In this case, campaigns do not have the chance to get many comrades, hence, the Opt-Out Modules may decide not to send opt-out requests because the number of comrades is not high enough. Both the Time Portal Attack and the Separation Attack are avoided by the way a Campaign Coordinator checks suitable start dates and times. The attacks are further mitigated by the possibility to participate in more than one campaign. 

\section{Conclusion and Future Work}\label{ch:conclusion}
The anti-spam approach presented in this paper increases the costs of websites advertised by spam and decreases their revenue, hence, making advertising by unsolicited email unattractive. Contrary to existing solutions, a decentralized approach using a DHT has been chosen to avoid attacks. The anti-spam approach offers protection against attacks like Man-in-the-Middle Attacks, Denial of Service Attacks, Time Portal Attacks as well as Separation Attacks. An implementation of the anti-spam solution for the Thunderbird mail client exists. Future work will target on the Opt-Out Module and the Target Evaluator.





%

\end{document}